\documentclass[aps,prb,twocolumn,groupedaddress,showpacs]{revtex4}

\usepackage{bm}
\usepackage{graphicx}

\begin{document}

\title{Bulk and surface low-energy excitations in YBa$_2$Cu$_3$O$_{7-\delta}$ studied by high-resolution angle-resolved photoemission spectroscopy}

\author{
	K. Nakayama,$^1$
	T. Sato,$^{1,2}$
	K. Terashima,$^1$
	H. Matsui,$^1$
	T. Takahashi,$^{1,2}$
	M. Kubota,$^{3}$
	K. Ono,$^{3}$
	T. Nishizaki,$^4$
	Y. Takahashi,$^4$
	and N. Kobayashi$^4$}

\affiliation{$^1$Department of Physics, Tohoku University, Sendai 980-8578, Japan}
\affiliation{$^2$CREST, Japan Science and Technology Agency (JST), Kawaguchi 332-0012, Japan}
\affiliation{$^3$Institute of Materials Structure Science, KEK, Tsukuba 305-0801, Japan}
\affiliation{$^4$Institute for Materials Research, Tohoku University, Sendai 980-8577, Japan}

\date{\today}

\begin{abstract}
We have performed high-resolution angle-resolved photoemission spectroscopy on YBa$_2$Cu$_3$O$_{7-\delta}$ (Y123; $\delta$ = 0.06; $T_{\rm c}$ = 92 K).  By accurately determining the Fermi surface and energy band dispersion, we solve long-standing controversial issues as to the anomalous electronic states of Y-based high-$T_{\rm c}$ cuprates.  We unambiguously identified surface-bilayer-derived bonding and antibonding bands, together with their bulk counterparts.  The surface bands are highly overdoped (hole concentration $x$ = 0.29), showing no evidence for the gap opening or the dispersion anomaly in the antinodal region, while the bulk bands show a clear $d_{x^2-y^2}$-wave superconducting gap and the Bogoliubov quasiparticle-like behavior with a characteristic energy scale of 50-60 meV indicative of a strong electron-boson coupling in the superconducting state.  All these results suggest that the metallic and superconducting states coexist at the adjacent bilayer of Y123 surface.
\end{abstract}

\maketitle

\section{INTRODUCTION}
Low-energy excitations play a key role in characterizing various physical properties such as superconductivity and metal-insulator transition.  In cuprate high-$T_{\rm c}$ superconductors (HTSCs), low-energy one-particle excitations are intensively studied with high-resolution angle-resolved photoemission spectroscopy (ARPES), which has a unique capability to directly observe the momentum-resolved electronic states.  Previous ARPES studies on Bi$_2$Sr$_2$CaCu$_2$O$_8$ (Bi2212) clarified several characteristic features essential to the anomalous superconductivity, such as a large Fermi surface (FS) centered at ($\pi$, $\pi$) point in the Brillouin zone (BZ),\cite{FSAebi,FSHong} the $d_{x^2-y^2}$-wave superconducting (SC) gap,\cite{GapHong,GapShen} the emergence of well-defined Bogoliubov quasiparticles below $T_{\rm c}$ in the antinodal region,\cite{FengScience,MatsuiBQP} and the pseudogap above $T_{\rm c}$ in the underdoped regime.\cite{HongPG,LoeserPG,NormanNature}  Recent remarkable progress in energy and momentum resolutions of ARPES further enables us to directly observe the bilayer splitting due to interaction of two CuO$_2$ planes\cite{FengBilayer,ChuangBilayer} and the strong mass renormalization of band (kink in the energy dispersion) in the vicinity of the Fermi level ($E_{\rm F}$) indicative of an electron-boson coupling.\cite{DamascelliReview,JCReview}  However, it is still unclear whether or not these characteristic low-energy excitations are generic feature for all HTSCs.  Hence the universality of essential character of CuO$_2$ plane should be carefully checked by performing ARPES with other families of HTSCs.  In this regard, YBa$_2$Cu$_3$O$_{7-\delta}$ (Y123) is the most suitable candidate since it is also a bilayered system with a similar maximum $T_{\rm c}$, and the bosonic (phononic and magnetic) excitations, which are related to the low-energy dispersion ($E$-$k$ relation), are intensively studied by inelastic neutron scattering experiments.\cite{INSReview}

It is known from previous ARPES on Y-based HTSCs \cite{Tobin,Gofron,Campuzano123,Schabel1,Schabel2} that (i) the SC gap is not as robust as that of Bi2212 and is hardly observed even below $T_{\rm c}$,\cite{Tobin} and (ii) strong surface states in the antinodal region dominate the low-energy excitations,\cite{Schabel1,Schabel2} causing a difficulty in distinguishing genuine feature of the bulk CuO$_2$ plane.  These problems have hindered for a long time further detailed investigations of the electronic states of Y-based HTSCs.  A recent ARPES study of Y123 by Lu $et$ $al$. \cite{Lu} distinguished the bulk and surface peaks and provided an evidence for the bulk SC-gap opening at X and Y points in BZ, as well as a characteristic ``peak-dip-hump" spectral line shape below $T_{\rm c}$.  Very recently, Borisenko $et$ $al$. \cite{Borisenko} reported the first evidence for the existence of kink in the dispersion along the nodal direction.  Although the experimental results from these two groups certainly provided a way to better understanding the anomalous electronic states of Y123, they also gave rise to serious contradictions/uncertainties in the assignment of band characters.  If both interpretations are correct, it follows that (i) the bilayer splitting observed along the nodal direction \cite{Borisenko} is absent in the antinodal region,\cite{Lu} and (ii) the surface state observed around X/Y points has nothing to do with the nodal dispersion.\cite{Borisenko}  Furthermore, there still remain several unresolved issues as to (i) the microscopic origin of the surface band, (ii) the $k$-dependence of the SC gap, and (iii) the origin of the electron-boson interaction responsible for the superconductivity.

In this article, we report high-resolution ARPES study on Y123 in order to address the above-mentioned issues.  By carefully tuning photon energy to separate contribution from the different bands, we have successfully mapped out the two-dimensional band dispersion and FS.  The results reveal existence of five different bands in the vicinity of $E_{\rm F}$, which were not well identified in previous studies.  They are the CuO-chain band,  the bonding and antibonding bands of heavily-overdoped crystal surface, and their bulk counterparts.  We also found that the bulk electronic states of CuO$_2$ plane certainly show a signature of strong electron-boson coupling, the $d_{x^2-y^2}$-wave SC gap, and relatively broad SC peak indicative of the existence of electronic inhomogeneity, showing a remarkable similarity with those of Bi2212.  Present result suggests the universality of bulk electronic states in bilayered HTSCs, and opens a way to the systematic ARPES investigations of Y-based HTSCs.

\section{EXPERIMENTS}
High-quality single crystals of untwinned and twinned nearly optimally-doped YBa$_2$Cu$_3$O$_{7-\delta}$ (Y123, $\delta$ = 0.06, $T_{\rm c}$ = 92 K) were grown by the self-flux method using yttria crucibles.  The oxygen concentration was controlled by annealing the samples under oxygen atmosphere at high temperature.  Details of sample preparation have been described elsewhere.\cite{NishizakiSample}  The $T_{\rm c}$ of samples was determined by the magnetic susceptibility measurement.  The hole concentration $x$ has been estimated to be 0.175.\cite{Liang}

ARPES measurements have been done using a VG-SCIENTA SES2002 spectrometer with a 5-axis manipulator (R-Dec, $i$ GONIO) \cite{Aiura} at a newly-developed BL28 beamline in Photon Factory (PF), KEK, Tsukuba.  Part of data has been obtained by SES2002 spectrometer at the undulator-4mNIM beamline in Synchrotron Radiation Center (SRC), Wisconsin.  We used circularly polarized lights at PF and linearly ones at SRC.  The energy and angular resolutions were set at 12-25 meV and 0.2$^\circ$, respectively.  The sample orientation was determined by the Laue x-ray diffraction pattern prior to ARPES measurements.  Clean surfaces for ARPES measurements were obtained by $in$ $situ$ cleaving of crystals in an ultrahigh vacuum of 1$\times$10$^{-10}$ Torr.  In order to avoid possible oxygen loss from the surface \cite{EdwardsSTM, Campuzano123}, we cleaved sample at 10 K.  Temperature of the sample has been kept at 10 K during the ARPES measurement, except for the temperature-dependent experiment.  The Fermi level ($E_{\rm F}$) of the sample was referenced to that of a gold film evaporated onto the sample substrate.

\section{RESULTS AND DISCUSSION}

\begin{figure}[!t]
\begin{center}
\includegraphics[width=3.0in]{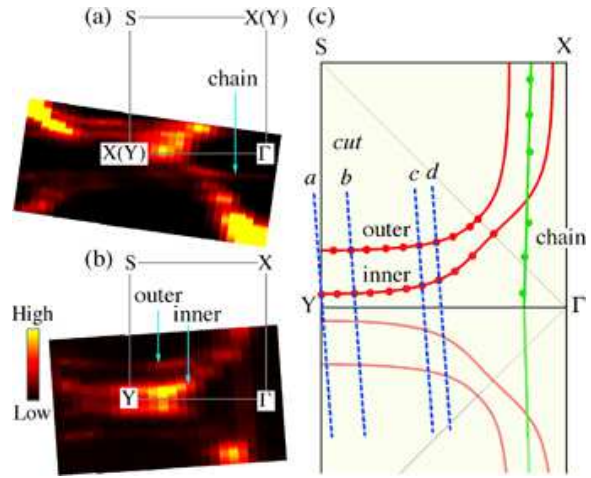}
\end{center}
\caption{
ARPES spectral intensity plots at $E_{\rm F}$ of Y123 ($T_{\rm c}$ = 92 K) as a function of two-dimensional wave vector measured at 10 K with 46-eV photons for (a) twinned and (b) untwinned samples.  ARPES intensity is integrated over the energy range of 30 meV centered at $E_{\rm F}$.  (c) Plot of experimentally-determined $k_{\rm F}$ points (solid circles).  Solid lines are the guide lines of FSs obtained by smoothly connecting determined $k_{\rm F}$ points.
}
\end{figure}

Figure 1 shows ARPES spectral intensity plots around Y(X) point in BZ measured at 10 K with 46-eV photons for twinned and untwinned Y123.  In both samples, we find two distinct intensity distributions arising from the CuO$_2$ plane.  The inner and outer FSs are nearly parallel around Y point, and do not merge even with approaching the nodal direction.  We also find additional straight FS segments coming from the CuO chain, while no signature of the chain FS is found around Y point in untwinned Y123 (Fig. 1(b)), in good agreement with previous reports.\cite{Tobin,Schabel1,Schabel2,Lu}  We do not find any evidence for the existence of a $c$(2$\times$2) shadow band as seen in Bi- and La-based HTSCs.\cite{FSAebi,Golden,Nakayama}  This is consistent with the previous ARPES report on Y123,\cite{Schabel2} supporting the structural origin of the shadow band.  By plotting the momentum location of local intensity maxima at $E_{\rm F}$ and tracing the crossing point of the energy band dispersion, we determined Fermi vectors ($k_{\rm F}$'s) for each band, and show the results in Fig. 1(c).  It is obvious that the observed large FS sheets of CuO$_2$ planes are holelike centered at S point with a marked difference in the volumes.

\begin{figure}[!t]
\begin{center}
\includegraphics[width=3.0in]{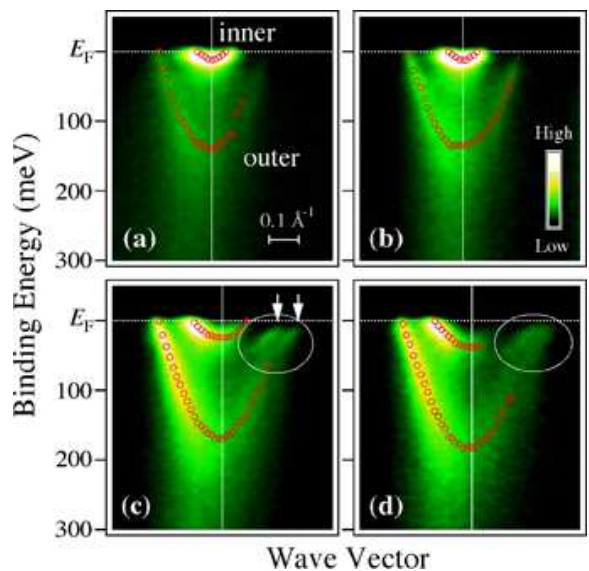}
\end{center}
\caption{
(a)-(d) ARPES spectral intensity plots of untwinned Y123 as a function of binding energy and wave vector measured along several cuts shown by dashed blue lines in Fig. 1(c).  Peak position of EDC for inner and outer bands after eliminating the effect from the Fermi-Dirac (FD) distribution function are shown by open red circles.  Area enclosed by white circles is the momentum region where the spectral intensity of the bulk bands is dominant.  White arrows represent the position of $k_{\rm F}$ points for bulk bands.
}
\end{figure}

Figure 2 shows ARPES spectral intensity plots of untwinned Y123 as a function of the binding energy and wave vector measured at four representative cuts shown in Fig. 1(c).  Along cut $a$ which corresponds to nearly YS direction, we observe highly-dispersive and weakly-dispersive bands having the bottom at 130 meV and 10 meV, respectively.  These bands show asymmetric intensity variation with respect to the $\Gamma$Y high symmetry line (white solid line), indicating presence of a strong matrix-element effect due to the circular polarization of incident light.  These bands disperse very weakly along $\Gamma$Y line around Y point (cut $b$).  In addition to these two bands, we unambiguously identify another feature near $E_{\rm F}$ in the momentum location where the spectral intensity of the outer and inner bands is significantly suppressed (area enclosed by circles).  As clearly visible in cut $c$, this feature consists of two independent bands, having smaller velocity as compared to the outer band, and showing a finite leading-edge shift at 10 K.  Similar feature is also observed in cut $d$ although the spectral intensity is broadly distributed as compared to that in cut $c$.

\begin{figure}[!t]
\begin{center}
\includegraphics[width=3.0in]{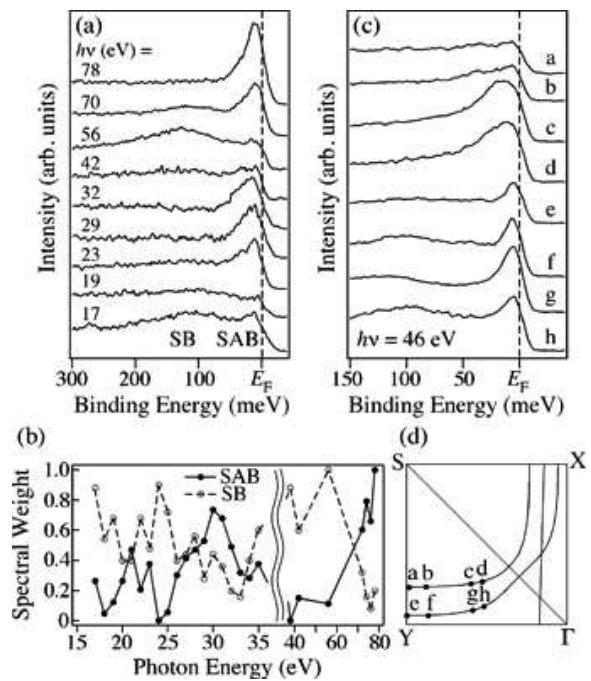}
\end{center}
\caption{
(a) Photon-energy dependence of ARPES spectrum at Y(X) point for twinned Y123.  ARPES spectra are normalized by the spectral intensity at 300 meV binding energy.  (b) Spectral weight of SB and SAB band as a function of $h\nu$, estimated by fitting the spectra in (a) by two asymmetric Lorentzians with a broad background\cite{MatsuiBQP} multiplied by the FD function.  The spectral weight of SB and SAB bands is normalized by the maximum weight at a specific $h\nu$ (56 eV and 78 eV for SB and SAB bands, respectively). (c) EDC at 10 K ($h\nu$ = 46 eV) measured at various $k_{\rm F}$ points of the surface bands shown in (d).
}
\end{figure}

In order to elucidate the character of inner and outer bands, we have performed photon-energy-dependent ARPES measurements at Y point.  The result is shown in Fig. 3(a).  The energy position of two bands does not show a discernible change as a function of $h\nu$, while the change in $h\nu$ value causes substantial difference in the intensity ratio of two bands.  Furthermore, the intensity weight of two bands, determined by the fitting of obtained energy distribution curves (EDC), shows an oscillating behavior as a function of $h\nu$, as illustrated in Fig. 3(b).  The oscillating behavior is understood by considering the existence/absence of available final states during the photoexcitation process.  It is also noted that each band shows the local maximum/minimum of spectral weight at different photon energies.  This trend, possibly caused by the difference in the symmetry of wave functions for two bands \cite{Lindroos}, is quite similar to the $h\nu$-dependent matrix-element effect of the bonding and antibonding bands in Bi2212,\cite{BorisenkoBilayer} demonstrating that the outer and inner bands are actually assigned as the bonding and antibonding bands, respectively.  It is noted here that previous assignment of a finite $k_{\rm z}$ dispersion in Y123 suggested by Schabel $et$ $al$. \cite{Schabel1} would be also influenced by this matrix-element effect.  This effect should be seriously taken into account when we discuss the $k_{\rm z}$ dispersion and the absolute intensity of bands in Y123, as in the case of Bi-based HTSCs.\cite{Bansil}

To check whether or not these bands show any signature of SC-gap opening, we plot in Fig. 3(c) EDC for various $k_{\rm F}$ points for both bands as shown in Fig. 3(d).  It is apparent that there is no observable leading-edge shift toward higher binding energy at any points on both FSs, showing that these bands are indeed metallic even well below bulk $T_{\rm c}$, indicating that these bands do not reflect the bulk superconductivity.  This is understood as a consequence of anomalous overdoping of CuO$_2$ planes, by considering an experimental fact that the hole concentration estimated from the average of the volume of the bonding and antibonding FSs (see Fig. 1(c)) is x = 0.29$\pm$0.02, which corresponds to the doping level at the boundary between the SC and metallic phases.\cite{ParabolaTc}   By also taking into account the facts that bulk superconducting properties are quite similar between Y123 and Bi2212,\cite{Uemura} and ARPES is relatively surface-sensitive experimental technique, it is plausible that the cleaved top-most CuO$_2$ bilayer is overdoped.  Therefore, we call the outer and the inner bands as the surface bonding (SB) and the surface antibonding (SAB) band, respectively.  Hence, previously assigned ``surface band" and ``hump feature" by Lu $et$ $al$. \cite{Lu} turn out to be the bilayer-split SAB and SB bands, respectively.  The size of bilayer splitting at Y point is estimated to be 120 meV, comparable to that observed in Bi2212 (90-110 meV).\cite{FengBilayer,ChuangBilayer,BorisenkoBilayer}  The gapped couple of bands in Fig. 2(c) would correspond to the bilayer-split bulk counterpart, so that the right-hand- and left-hand-side bands, as indicated by white arrows, are assigned as the bulk bonding (BB) and the bulk antibonding (BAB) bands, respectively.  Since we have successfully distinguished the bulk and surface bands, the character of these bands can be now evaluated in more detail.  Especially, the self-energy analysis of SB and SAB bands provides a good chance to quantitatively elucidate the nature of quasiparticle dynamics of CuO$_2$ plane in a heavily-overdoped limit.

\begin{figure}[!t]
\begin{center}
\includegraphics[width=3.0in]{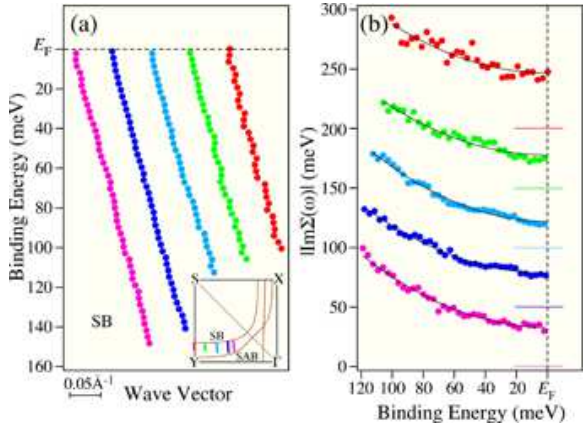}
\end{center}
\caption{
(a) Band dispersion near $E_{\rm F}$ of the SB band determined by fitting the MDC along several cuts shown by solid lines in the inset.  (b) Imaginary part of the self energy $|$Im$\Sigma$($\omega$)$|$ obtained by fitting the MDC.  For better illustration, the energy dispersion and $|$Im$\Sigma$($\omega$)$|$ are shown by adding offsets.  Solid lines near $E_{\rm F}$ in (b) correspond to the zero point of each $|$Im$\Sigma$($\omega$)$|$.  Black curves are the result of fitting by using $|$Im$\Sigma$($\omega$)$|$ = $\alpha\omega^2$+$\beta$.
}
\end{figure}

Figure 4(a) shows the energy band dispersion near $E_{\rm F}$ of the SB band for several representative cuts determined by fitting the momentum distribution curves  (MDC).  We used ARPES data at left-half side with respect to $\Gamma$Y line in Figs. 2(a)-(d), since the surface states dominate the ARPES intensity in this momentum region.  It is noted that the bottom of SAB band is located very close to $E_{\rm F}$ so that the MDC analysis up to higher energy is not applicable.  As seen in Fig. 4(a), we find a fairly straight dispersion of the SB band along all the momentum cuts around the antinodal region (see inset) with no signature of the dispersion kink within the experimental accuracy.  The absence of kink in Fig. 4(a) is also consistent with the plots of $|$Im$\Sigma$($\omega$)$|$ in Fig. 4(b) where there is no indication of a sudden drop at a characteristic energy scale as seen in previous ARPES reports of other cuprates.\cite{DamascelliReview,JCReview,AdamQP,Johnson,ImSigmaLSCO}  As also seen in Fig. 4(b), $|$Im$\Sigma$($\omega$)$|$ does not show the $\omega$-linear behavior unlike optimally-doped Bi2212 above $T_{\rm c}$ (ref. 34), but shows a quadratic behavior as seen in the transport and previous ARPES measurements of heavily-overdoped cuprates.\cite{ODtransport,ODARPES}  These experimental results also support the overdoping of crystal surface in Y123.

\begin{figure}[!t]
\begin{center}
\includegraphics[width=3.0in]{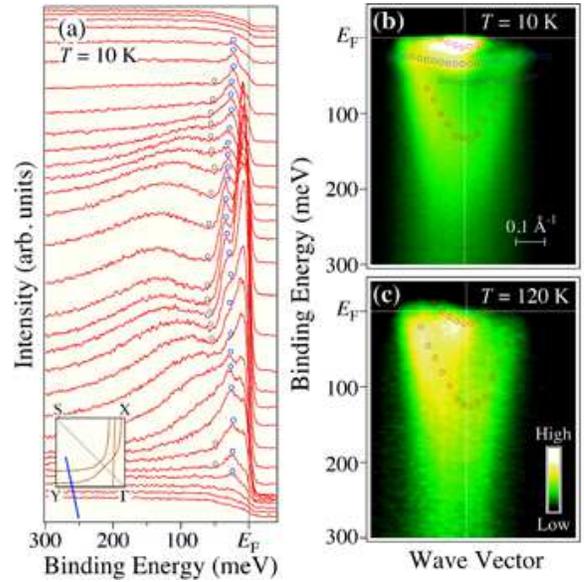}
\end{center}
\caption{
(a) EDC and (b) its intensity plot at 10 K for untwinned Y123 as a function of binding energy and wave vector, and (c) the intensity plot at 120 K, measured at a cut shown by blue line in inset to (a).  Peak position of EDC for surface and bulk bands, as well as dip (break) are indicated by red, blue, and gray open circles, respectively.  Location of $\Gamma$Y-line is indicated by thin solid lines in (b) and (c).
}
\end{figure}

Since the surface states are well characterized, we now discuss the properties of bulk bands.  Figure 5 shows a set of EDC measured at 10 K along a cut around the antinodal region shown in the inset, together with the intensity plot.  To see more clearly the bulk band, we slightly changed the polarization direction of the incident light with respect to the sample by rotating the sample azimuth, since the rotation of sample even by a few degrees was found to enhance significantly the sensitivity of the bulk band.  As seen in Fig. 5(a), we observe bulk peak (open blue circles) in addition to the SB and SAB bands, although BB and BAB components are indistinguishable because two peaks are close and may overlap in this momentum region.  This bulk peak has a bottom at $\Gamma$Y line, and gradually approaches $E_{\rm F}$ but never crosses $E_{\rm F}$, indicative of SC-gap opening.  As seen in Fig. 5(c), the bulk peak disappears above $T_{\rm c}$ (120 K), demonstrating that it is assigned as a Bogoliubov quasiparticle peak.  Another interesting feature in Fig. 5(a) is that a dip or a spectral break emerges at higher energy than the SC peak, as indicated by the gray open circles.  This dip/break originates in a nearly optimally-doped bulk component, since it is related to the bulk SC peak.  The existence of the spectral dip/break suggests that a strong electron-boson coupling exists in bulk Y123 around the optimally-doped region.  The similarity of spectral line shape between Y123 and Bi2212\cite{AdamKdep} suggests that the origin of the electron-boson coupling in these two types of bilayered cuprates is understood with the common framework.  The spectral dip in Y123 is observed at the characteristic energy scale of 50-60 meV, slightly lower than the dip and the energy position of the kink in Bi2212 (70 meV).  The finite difference in their energy scales by 10-20 meV may be explained by the difference in the size of the SC gap.  It is necessary to determine the momentum- and temperature-dependence of the dispersion kink in detail to elucidate the origin of the electron-boson coupling.

To elucidate the $k$-dependence of the SC gap, we at first determined the $k_{\rm F}$ position of bulk bands at various $k_{\rm F}$ points, by employing the minimum-gap locus method.\cite{NormanNature}  The result shown in Fig. 6(a) indicates that the $k_{\rm F}$ points of the BB and BAB bands are well outside the $k_{\rm F}$ points of SAB band, while they are nearly on the SB band at the off-nodal region.  This demonstrates that the bulk bands are much less doped than the surface bands.  Next, we plot in Fig. 6(b) EDC for untwinned Y123 at 10 K measured at various $k_{\rm F}$ points of bulk bands characterized by FS angle ($\phi$) defined in Fig. 6(a).  We do not plot the data around X point to avoid complication from the chain band.  As seen in Fig. 6(b), the metallic nature of the surface state is remarkably suppressed.  Although there is a small but finite surface contribution around $E_{\rm F}$, this does not seriously affect the determination of the size of bulk SC gap, since the bulk SC peak is well identified in wide area of BZ.  It is apparent from the figure that energy position of the bulk SC peak is strongly $k$-dependent and gradually becomes small on approaching the nodal direction, indicating the $d_{x^2-y^2}$-like order parameter.  In order to estimate the SC gap size ($\Delta$), we have symmetrized the spectra with respect to $E_{\rm F}$ (ref. 9) as shown in Fig. 6(c) and fit each curve by two Lorentzians symmetric with respect to $E_{\rm F}$.  The obtained $k$-dependence of $\Delta$ is shown in Fig. 6(d).  Apparently, the gap become small with approaching $\phi$ = 45$^{\circ}$, showing the $d_{x^2-y^2}$-like nature.\cite{notenode}  We have fit the gap size as a function of $\phi$ by using the $d_{x^2-y^2}$-wave gap function ($\Delta$($\phi$) = ${\Delta}_{max}$cos(2$\phi$)).  A good agreement is obtained with the parameter ${\Delta}_{max}$ = 34 meV.  This value is similar to the SC peak position in the tunneling measurement.\cite{EdwardsSTM}  On the other hand, this value is smaller than that estimated by Lu $et$ $al$. (44 meV) \cite{Lu} because they measured the SC gap at Y point but not at the $k_{\rm F}$ point.  It is noted here that observed energy width of the SC quasiparticle peak ($\sim$20 meV) is not resolution limited (energy resolution: 12 meV).  A similar behavior has been reported in Bi2212 \cite{Ding} and explained in terms of the existence of spatially inhomogeneous electronic states as observed by STM/STS experiments on Bi2212.\cite{Pan}  When the SC-gap size shows spatial variation, the spectral peak is expected to be broadened, since ARPES probes photoelectrons emitted from a large area of the sample surface ($\sim$100 $\mu$m).  This implies that, even in Y123 which is reported to be homogeneous on the scale of coherence length ($\sim$100 nm)\cite{Yeh}, the inhomogeneity certainly affects the electronic states on the wider scale ($\sim$100 $\mu$m).  This conclusion is supported by the result of previous NQR study \cite{Ofer} which suggests the existence of charge inhomogeneity in Y123.

\begin{figure}[!t]
\begin{center}
\includegraphics[width=3.0in]{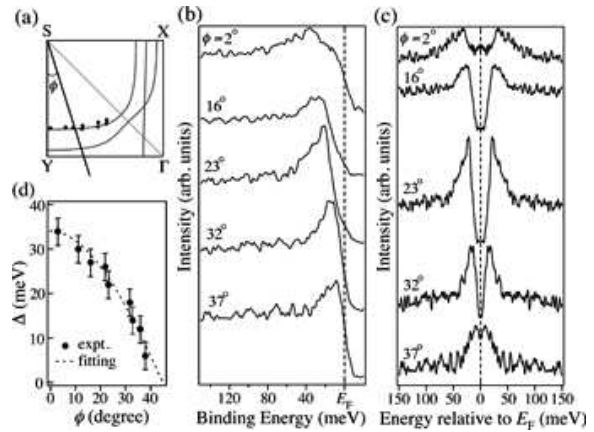}
\end{center}
\caption{
(a) Location of $k_{\rm F}$ points of the BB and BAB bands (solid circles) together with the definition of the FS angle ($\phi$).  (b) ARPES spectra at 10 K of untwinned Y123 measured at various $k_{\rm F}$ points of the bulk bands shown in (a).  (c) Symmetrized ARPES spectra of (b).  (d) $k$-dependence of the superconducting gap ($\Delta$) as a function of $\phi$.  Dashed line represents the best fit using the $d_{x^2-y^2}$-wave gap function.
}
\end{figure}

Now we discuss the origin of surface state.  We conclude that the surface bands are highly overdoped as compared to the bulk from the following experimental results;  (i) there are no gap opening well below bulk $T_{\rm c}$ (Fig. 3(c)), (ii) there is no evidence for the dispersion kink near $E_{\rm F}$ (Fig. 4(a)), (iii) $|$Im$\Sigma$($\omega$)$|$ shows $\omega^2$-like energy dependence (Fig. 4(b)) similar to other overdoped cuprates, and (iv) the estimated hole concentration is very large (x = 0.29).  This anomalous overdoping would be characteristic of the cleaved surface of Y123.  It is established from previous STM measurements \cite{EdwardsSTM,NishizakiJPSJ} that Y123 cleaves between the BaO layer and the CuO chain.  As a result, two scenarios are possible to explain the mechanism of surface overdoping.  First is a overdoping of the CuO-terminated surface, and the second is the overdoping of BaO-terminated surface.   In Y123, the CuO chain provides holes to the CuO$_2$ sheets located at both side of the chain in the bulk crystal.  In the first case, it is expected that the excess holes are provided to the residual CuO$_2$ sheets, since one side of the CuO$_2$ sheets is completely removed by the cleaving.  The second case is consistent with the STM/STS results where the BaO-terminated surface shows metallic behavior with no SC-gap opening,\cite{EdwardsSTM,NishizakiJPSJ} although the microscopic mechanism for the overdoping is unclear.  By also taking into account the fact that the SC gap has been observed only on the CuO-terminated surface \cite{EdwardsSTM,NishizakiJPSJ}, the superconductivity may disappear at the CuO$_2$ sheets just below BaO layer due to the absence of the CuO chain, while the superconductivity may be maintained at the sheets below the CuO chain.  In either case, it is suggested that cleavage strongly influences the carrier distribution of the crystal surface and causes the overdoping of surface CuO$_2$ layers.  To elucidate the character of surface electronic states in more detail, further comprehensive high-resolution ARPES and STM/STS studies are necessary.
				
\section{CONCLUSION}
We reported high-resolution ARPES results on Y123 to elucidate the character of low-energy excitations.  We clarified that the electronic states of Y123 near $E_{\rm F}$ consist of five different bands, the CuO chain band, the surface bonding and antibonding bands, and the bulk bonding and antibonding bands.  We revealed that the surface bands are highly overdoped by the cleaving.  On the other hand, the bulk bands show a clear sensitivity to the SC transition and a signature of electron-boson coupling with a characteristic energy scale of 50-60 meV.  These experimental results unambiguously provide a convincing explanation on the puzzling and anomalous electronic states of Y123.

\acknowledgments
{
This work is supported by grants from JST-CREST and MEXT of Japan.  K. T. thanks JSPS for financial support.
}


\begin{references}
\bibitem{FSAebi} P. Aebi, J. Osterwalder, P. Schwaller, L. Schlapbach, M. Shimoda, T. Mochiku, and K. Kadowaki, Phys. Rev. Lett. {\boldmath $72$}, 2757 (1994).
\bibitem{FSHong} H. Ding, A. F. Bellman, J. C. Campuzano, M. Randeria, M. R. Norman, T. Yokoya, T. Takahashi, H. Katayama-Yoshida, T. Mochiku, K. Kadowaki, G. Jennings, and G. P. Brivio, Phys. Rev. Lett. {\boldmath $76$}, 1533 (1996).
\bibitem{GapHong} H. Ding, M. R. Norman, J. C. Campuzano, M. Randeria, A. F. Bellman, T. Yokoya, T. Takahashi, T. Mochiku, and K. Kadowaki, Phys. Rev. B {\boldmath $54$}, R9678 (1996).
\bibitem{GapShen} Z.-X. Shen, D. S. Dessau, B. O. Wells, D. M. King, W. E. Spicer, A. J. Arko, D. Marshall, L. W. Lombardo, A. Kapitulnik, P. Dickinson, S. Doniach, J. DiCarlo, A. G. Loeser, and C. H. Park, Phys. Rev. Lett. {\boldmath $70$}, 1553 (1993).
\bibitem{FengScience} D. L. Feng, D. H. Lu, K. M. Shen, C. Kim, H. Eisaki, A. Damascelli, R. Yoshizaki, J.-i. Shimoyama, K. Kishio, G. D. Gu, S. Oh, A. Andrus, J. O'Donnell, J. N. Eckstein, and Z.-X. Shen, Science {\boldmath $289$}, 277 (2000).
\bibitem{MatsuiBQP} H. Matsui, T. Sato, T. Takahashi, S.-C. Wang, H.-B. Yang, H. Ding, T. Fujii, T. Watanabe, and A. Matsuda, Phys. Rev. Lett. {\boldmath $90$}, 217002 (2003).
\bibitem{HongPG} H. Ding, T. Yokoya, J. C. Campuzano, T. Takahashi, M. Randeria, M. R. Norman, T. Mochiku, K. Kadowaki, and J. Giapintzakis, Nature (London) {\boldmath $382$}, 51 (1996).
\bibitem{LoeserPG} A. G. Loeser, Z.-X. Shen, D. S. Dessau, D. S. Marshall, C. H. Park, P. Fournier, and A. Kapitulnik, Science {\boldmath $273$}, 325 (1996).
\bibitem{NormanNature} M. R. Norman, H. Ding, M. Randeria, J. C. Campuzano, T. Yokoya, T. Takeuchi, T. Takahashi, T. Mochiku, K. Kadowaki, P. Guptasarma, and D. G. Hinks, Nature (London) {\boldmath $392$}, 157 (1998).
\bibitem{FengBilayer} D. L. Feng, N. P. Armitage, D. H. Lu, A. Damascelli, J. P. Hu, P. Bogdanov, A. Lanzara, F. Ronning, K. M. Shen, H. Eisaki, C. Kim, J.-i. Shimoyama, K. Kishio, and Z.-X. Shen, Phys. Rev. Lett. {\boldmath $86$}, 5550 (2001).
\bibitem{ChuangBilayer} Y.-D. Chuang, A. D. Gromko, A. Fedorov, Y. Aiura, K. Oka, Yoichi Ando, H. Eisaki, S. I. Uchida, and D. S. Dessau, Phys. Rev. Lett. {\boldmath $87$}, 117002 (2001).
\bibitem{DamascelliReview} A. Damascelli, Z. Hussain, and Z.-X. Shen, Rev. Mod. Phys. {\boldmath $75$}, 473 (2003).
\bibitem{JCReview} J. C. Campuzano, M. R. Norman, and M. Randeria, {\it The Physics of Superconductors}, edt. K. H. Bennemann, J. B. Bennemann, and J. B. Ketterson (Springer, New York, 2003).
\bibitem{INSReview} P. Bourges, {\it The Gap Symmetry and Fluctuations in High Temperature Superconductors}, edt. J. Bok, G. Deutscher, D. Pavuna, and S. A. Wolf, (Plenum Press, New York, 1998).
\bibitem{Tobin} J. G. Tobin, C. G. Olson, C. Gu, J. Z. Liu, F. R. Solal, M. J. Fluss, R. H. Howell, J. C. O'Brien, H. B. Radousky, and P. A. Sterne, Phys. Rev. B {\boldmath $45$}, 5563 (1992).
\bibitem{Gofron} K. Gofron, J. C. Campuzano, A. A. Abrikosov, M. Lindroos, A. Bansil, H. Ding, D. Koelling, and B. Dabrowski, Phys. Rev. Lett. {\boldmath $73$}, 3302 (1994).
\bibitem{Campuzano123} J. C. Campuzano, G. Jennings, M. Faiz, L. Beaulaigue, B. W. Veal, J. Z. Liu, A. P. Paulikas, K. Vandervoort, H. Claus, R. S. List, A. J. Arko, and R. J. Bartlett, Phys. Rev. Lett. {\boldmath $64$}, 2308 (1990).
\bibitem{Schabel1} M. C. Schabel, C.-H. Park, A. Matsuura, Z.-X. Shen, D. A. Bonn, R. Liang, and W. N. Hardy, Phys. Rev. B {\boldmath $57$}, 6090 (1998).
\bibitem{Schabel2} M. C. Schabel, C.-H. Park, A. Matsuura, Z.-X. Shen, D. A. Bonn, R. Liang, and W. N. Hardy, Phys. Rev. B {\boldmath $57$}, 6107 (1998).
\bibitem{Lu} D. H. Lu, D. L. Feng, N. P. Armitage, K. M. Shen, A. Damascelli, C. Kim, F. Ronning, Z.-X. Shen, D. A. Bonn, R. Liang, W. N. Hardy, A. I. Rykov, and S. Tajima, Phys. Rev. Lett. {\boldmath $86$}, 4370 (2001).
\bibitem{Borisenko} S. V. Borisenko, A. A. Kordyuk, V. Zabolotnyy, J. Geck, D. Inosov, A. Koitzsch, J. Fink, M. Knupfer, B. Buchner, V. Hinkov, C. T. Lin, B. Keimer, T. Wolf, S. G. Chiuzb\"aian, L. Patthey, and R. Follath, Phys. Rev. Lett. {\boldmath $96$}, 117004 (2006).
\bibitem{NishizakiSample} T. Nishizaki, K. Shibata, T. Naito, M. Maki, and N. Kobayashi, J. Low Temp. Phys. {\boldmath $117$}, 1375 (1999).
\bibitem{Liang} R. Liang, D. A. Bonn, and W. N. Hardy, Phys. Rev. B {\boldmath $73$}, 180505(R) (2006).
\bibitem{Aiura} Y. Aiura, H. Bando, T. Miyamoto, A. Chiba, R. Kitagawa, S. Maruyama, and Y. Nishihara, Rev. Sci. Instrum. {\boldmath $74$}, 3177 (2003).
\bibitem{EdwardsSTM} H. L. Edwards, J. T. Markert, and A. L. de Lozanne, Phys. Rev. Lett. {\boldmath $69$}, 2967 (1992).
\bibitem{Golden} A. Mans, I. Santoso, Y. Huang, W. K. Siu, S. Tavaddod, V. Arpiainen, M. Lindroos, H. Berger, V. N. Strocov, M. Shi, L. Patthey, and M. S. Golden, Phys. Rev. Lett. {\boldmath $96$}, 107007 (2006).
\bibitem{Nakayama} K. Nakayama, T. Sato, T. Dobashi, K. Terashima, S. Souma, H. Matsui,	T. Takahashi, J. C. Campuzano, K. Kudo, T. Sasaki, N. Kobayashi, T. Kondo, T. Takeuchi, K. Kadowaki, M. Kofu, and K. Hirota, Phys. Rev. B {\boldmath $74$}, 054505 (2006).
\bibitem{Lindroos} M. Lindroos, S. Sahrakorpi, and A. Bansil, Phys. Rev. B {\boldmath $65$}, 054514 (2002).
\bibitem{BorisenkoBilayer} A. A. Kordyuk, S. V. Borisenko, T. K. Kim, K. A. Nenkov, M. Knupfer, J. Fink, M. S. Golden, H. Berger, and R. Follath, Phys. Rev. Lett. {\boldmath $89$}, 077003 (2002).
\bibitem{Bansil} A. Bansil and M. Lindroos, Phys. Rev. Lett. {\boldmath $83$}, 5154 (1999).
\bibitem{ParabolaTc} M. R. Presland, J. L. Tallon, R. G. Buckley, R. S. Liu, and N. R. Flower, Physica C {\boldmath $176$}, 95 (1991).
\bibitem{Uemura} Y. J. Uemura, G. M. Luke, B. J. Sternlieb, J. H. Brewer, J. F. Carolan, W. N. Hardy, R. Kadono, J. R. Kempton, R. F. Kiefl, S. R. Kreitzman, P. Mulhern, T. M. Riseman, D. Ll. Williams, B. X. Yang, S. Uchida, H. Takagi, J. Gopalakrishnan, A. W. Sleight, M. A. Subramanian, C. L. Chien, M. Z. Cieplak, Gang Xiao, V. Y. Lee, B. W. Statt, C. E. Stronach, W. J. Kossler, and X. H. Yu, Phys. Rev. Lett. {\boldmath $62$}, 2317 (1989).
\bibitem{AdamQP} A. Kaminski, J. Mesot, H. Fretwell, J. C. Campuzano, M. R. Norman, M. Randeria, H. Ding, T. Sato, T. Takahashi, T. Mochiku, K. Kadowaki, and H. H\"ochst, Phys. Rev. Lett. {\boldmath $84$}, 1788 (2000).
\bibitem{Johnson} P. D. Johnson, T. Valla, A. V. Fedorov, Z. Yusof, B. O. Wells, Q. Li, A. R. Moodenbaugh, G. D. Gu, N. Koshizuka, C. Kendziora, Sha Jian, and D. G. Hinks, Phys. Rev. Lett. {\boldmath $87$} (2001) 177007.
\bibitem{ImSigmaLSCO} X. J. Zhou, Junren Shi, T. Yoshida, T. Cuk, W. L. Yang, V. Brouet, J. Nakamura, N. Mannella, Seiki Komiya, Yoichi Ando, F. Zhou, W. X. Ti, J. W. Xiong, Z. X. Zhao, T. Sasagawa, T. Kakeshita, H. Eisaki, S. Uchida, A. Fujimori, Zhenyu Zhang, E. W. Plummer, R. B. Laughlin, Z. Hussain, and Z.-X. Shen, Phys. Rev. Lett. {\boldmath $95$}, 117001 (2005).
\bibitem{ODtransport} Y. Kubo, Y. Shimakawa, T. Manako, and H. Igarashi, Phys. Rev. B {\boldmath $43$}, 7875 (1991).
\bibitem{ODARPES} A. Koitzsch, S. V. Borisenko, A. A. Kordyuk, T. K. Kim, M. Knupfer, J. Fink, H. Berger, and R. Follath, Phys Rev. B {\boldmath $69$}, 140507(R) (2004).
\bibitem{AdamKdep} A. Kaminski, M. Randeria, J. C. Campuzano, M. R. Norman, H. Fretwell, J. Mesot, T. Sato, T. Takahashi, and K. Kadowaki, Phys. Rev. Lett. {\boldmath $86$}, 1070 (2001).
\bibitem{notenode} It is difficult to determine the SC gap size of bulk bands along the nodal cut ($\phi$ = 45$^\circ$) because the bulk and surface bands are indistinguishable.
\bibitem{Ding} H. Ding, J. R. Engelbrecht, Z. Wang, J. C. Campuzano, S.-C. Wang, H.-B. Yang, R. Rogan, T. Takahashi, K. Kadowaki, and D. G. Hinks, Phys. Rev. Lett. {\boldmath $87$}, 227001 (2001).
\bibitem{Pan} S. H. Pan, J. P. O'Neal, R. L. Badzey, C. Chamon, H. Ding, J. R. Engelbrecht, Z. Wang, H. Eisaki, S. Uchida, A. K. Gupta, K.-W. Ng, E. W. Hudson, K. M. Lang, and J. C. Davis, Nature {\boldmath $413$}, 282 (2001).
\bibitem{Yeh} N.-C. Yeh, C.-T. Chen, G. Hammerl, J. Mannhart, A. Schmehl, C. W. Schneider, R. R. Schulz, S. Tajima, K. Yoshida, D. Garrigus, and M. Strasik, Phys. Rev. Lett. {\boldmath $87$}, 087003 (2001).
\bibitem{Ofer} R. Ofer, S. Levy, A. Kanigel, and A. Keren, Phys. Rev. B {\boldmath $73$}, 012503 (2006).
\bibitem{NishizakiJPSJ} M. Maki, T. Nishizaki, K. Shibata, and N. Kobayashi, J. Phys. Soc. Jpn. {\boldmath $70$}, 1877 (2001).

\end{references}
\end{document}